# SPHEREx Discovery of Strong Water Ice Absorption and an Extended Carbon Dioxide Coma in 3I/ATLAS

C.M. Lisse, Y.P. Bach, S. Bryan, B.P. Crill, A. Cukierman, O. Doré, B. Fabinsky, A. Faisst, P. M. Korngut, G. Melnick, Z. Rustamkulov, V. Tolls, M. Werner, M.L. Sitko, C. Champagne, M. Connelley, J.P. Emery, Y.R. Fernandez, B. Yang, and the SPHEREx Science Team

**Abstract.** In mid-August 2025, 0.75-5.0μm SPHEREx imaging spectrophotometric and ancillary NASA-IRTF/SpeX 0.7-2.5μm low-resolution spectral observations of 3I/ATLAS were obtained. The combined spectrophotometry is dominated by features due to water ice absorption and $CO_2$ gas emission. A bright, ~3' radius $CO_2$ gas coma was clearly resolved, corresponding to $Q_{gas,CO2}$~$9.4 \times 10^{26}$ molecules/sec. From the SPHEREx photometry, we put conservative, preliminary 3σ upper limits on the gas production rates for $H_2O$ and CO of $1.5 \times 10^{26}$ and $2.8 \times 10^{26}$ molecules/sec, respectively. No obvious jet, tail, or trail structures were found in SPHEREx images. If we assume all observed 1μm flux is scattered light from a $p_v$ = 0.04 albedo spherical nucleus, then the radius would be $R_{nuc}$~23km. Given the nucleus size limit $R_{nuc}$ < 2.8km from Jewett+ 2025, we conclude >99% of the measured SPHEREx continuum flux is from coma dust.

**1. SPHEREx Observations.** From 08-Aug- to 12-Aug-2025, the NASA SPHEREx mission (Doré+2018, Crill+2020) observed interstellar object 3I/ATLAS (hereafter 3I) from 0.75-5.0μm at the heliocentric distances 3.1-3.3au. The observations consisted of 72 special pointings of 115 seconds each, with 3I centered on unique positions across SPHEREx's 6 LVF-patterned detectors. Approximately 2" of object motion occurred during each integration, smaller than a SPHEREx 6.15" x 6.15" pixel. The image position of 3I was determined to <1" using the JPL Horizons Orbit #22, the mission software packages *skyloc* and *kete*, plus the SPHEREx Sky Simulator (Crill+2025). Photometry was obtained by locating 3I, placing a 2-pixel radius aperture at 3I's location, and subtracting the background counts from a surrounding annulus. While the SPHEREx platescale was optimal for detecting 3I's extended morphology and gas emission, a significant fraction of the pointings were contaminated by flux from background objects (galactic latitude$_{3I}$ was ~23°) and removed from further analysis until the background sky, without 3I, is re-observed in Nov-Dec 2025. For the reflectance spectra, the photometry was corrected for 3I's motion versus the Sun and Earth assuming a $1/r_h^2 * 1/\Delta^2$ dependence. No correction for 3I's phase function, rotation, or production rate variability was applied.

**2. Ancillary Measurements.** On 11-Aug-2025, 3I was observed from the NASA/IRTF 3.2m using the SpeX imaging spectrometer (Rayner+2003) with 0.7-2.5μm, R~200 spectroscopy. Two G2V stars (HD148744 and SA110-361) were used as local and absolute solar analogue references, respectively.

**3. Spectral Results**. The combined SPHEREx and IRTF photometry/reflectance spectrophotometry is shown in Figure 1a. There is good consistency between the different measurements and published results, showing a generally rising slope from 0.7-1.0μm and a flattened slope from 1-1.5μm. Clear IRTF/SpeX absorptions at 1.5, 2.1, and 2.4μm are new, although the latter is in a region of low SNR. SPHEREx confirms the downward trend at the end of the IRTF/SpeX spectrum, and continues its very strong drop through 2.5-3.1μm not readily observable from the ground for this V~16.6 magnitude object. The SPHEREx reflectance drops to ~20% of its 1μm flux level by 3μm, and stays low out to 5μm except for a strong peak located at ~4.3μm. The combined reflectance spectral shape is consistent with the $CO_2/H_2O$ ice+organics KBO "Cliff" category of Pinilla-Alonso+2025 (Figure 1a).

**4. Imaging Results.** A SPHEREx continuum dust image of 3I created by stacking 1.0-1.5μm images is shown in Figure 1b. 3I looks like a point source. SPHEREx images taken near the $H_2O$ (2.8-3.1μm), $CO_2$ (4.25-4.2μm), and CO (4.7μm) gas emission lines are shown in Figures 1c-e. While no obvious coma was found in $H_2O$ or CO, the $CO_2$ image shows a central condensation and detectable structure above the background with radius ≳3', far beyond the central PSF (Figure 1f). The observed $CO_2$ coma is roughly symmetric with respect to the Sun and orbital velocity directions.

**5. Temporal Variability**. Continuum SPHEREx geometry-corrected flux measurements of 3I at 1.0-1.5μm were corrected for the Solar spectrum and plotted against time. We see no evidence for variability of more than 20% over the timespan 08–12 Aug 2025, giving us confidence in the construction of our preliminary 0.75-5.0μm spectrum.

**6. Discussion/Interpretation.** Adopting $N_{molec} = 4\pi r_h^2 \Delta^2 F_v d(\lambda)/(g*h*\lambda)$ (Lisse+2009), and $Q_{gas} = N_{gas} * (2\pi v_{gas} \rho)$, where $r_h = 3.2$au, $\Delta = 2.6$au, the fluorescence efficiency $g$ = 2.9e-4, 2.6e-3, and 2.6e-4 for $H_2O$ 3.05μm, $CO_2$ 4.25μm, and CO 4.7μm emission, respectively (Crovisier+1997), $N_{gas}$ is the column number density of gas, $v_{gas}$ is the gas emission velocity 0.85/sqrt(3.2au) = 0.48 km/sec (Cochran+2012), and ρ is the projected distance on the sky (~11,600 km/pixel), we find the estimated rate of $CO_2$ gas loss, $Q_{gas,CO2}$~$9.4\times10^{26}$ ± 30% molecules/sec. Conservative, preliminary 3-σ upper limits for $Q_{gas,H2O}$ = $1.5\times10^{26}$ and for $Q_{gas,CO}$ = $2.8\times10^{26}$ molecules/sec were also found. This rate of $CO_2$ emission and upper limit for CO production is consistent with the activity of thermally processed short-period Solar System comets at 3.2au (Harrington-Pinto & Womack 2022).

From extant surveys, comets are known to be mixtures of ice and refractory dust, with $H_2O$, $CO_2$, and CO ices dominating the icy component (Bockelee-Morvan & Biver 2017). 3I appears to be an active object with morphology created mainly by $CO_2$ gas emission effects. This is physically reasonable, since 3I at $r_h$~3.2au was well within the Solar system's $CO_2$ "ice line" at ~14au during the observations (Lisse+2021, 2022). The strong reflectance signature of water ice can be explained by the presence of abundant water ice in 3I's nucleus and coma dust particles. The lack of a bright water gas coma is puzzling as 3I was not far outside the Solar system's "water ice line" at 2.5au during the observations. Most likely, 3I is emitting large chunks of mixed $CO_2$+$H_2O$ ice into its coma, where evaporative cooling of $CO_2$ is pinning the chunks' temperature at ~120K and greatly suppressing the $H_2O$ ice's vapor pressure (Lisse+2021).

The lack of any obvious dust coma extension at the 6"-scale is consistent with the ~3"-wide by 5"-long dust continuum morphology reported by Jewitt+2025. Assuming all observed 1-μm flux is scattered light from the nucleus and an $p_v$ = 0.04 albedo, ϕ(a)= 0.035 mag/deg surface, the radius $R_{nuc}$ of a spherical nucleus emitting $F_{scat}$ = 1.5±0.05 mJy = ($p_v *10^{-0.4*\phi(\alpha)} * \pi\ R_{nuc}^2$) ($F_{sun}/\pi$)/($r_h^2 \Delta^2$) (Lisse+1999) at λ=1μm is 23±0.7 km. This is ~10 times the nucleus size limit found by Jewitt+2025 using HST imaging, arguing that >99% of the observed SPHEREx flux is due to scattering from icy coma dust.

**7. Future Work.** The photometric measurements reported here are from the first ½ of all the August 2025 SPHEREx 3I pointings. Treatment of stellar contamination will greatly improve after the spacecraft re-observes the same field in 6 months. 3I/ATLAS will also pass through SPHEREx's planned survey pattern again in November-December 2025.

# 9. Figure

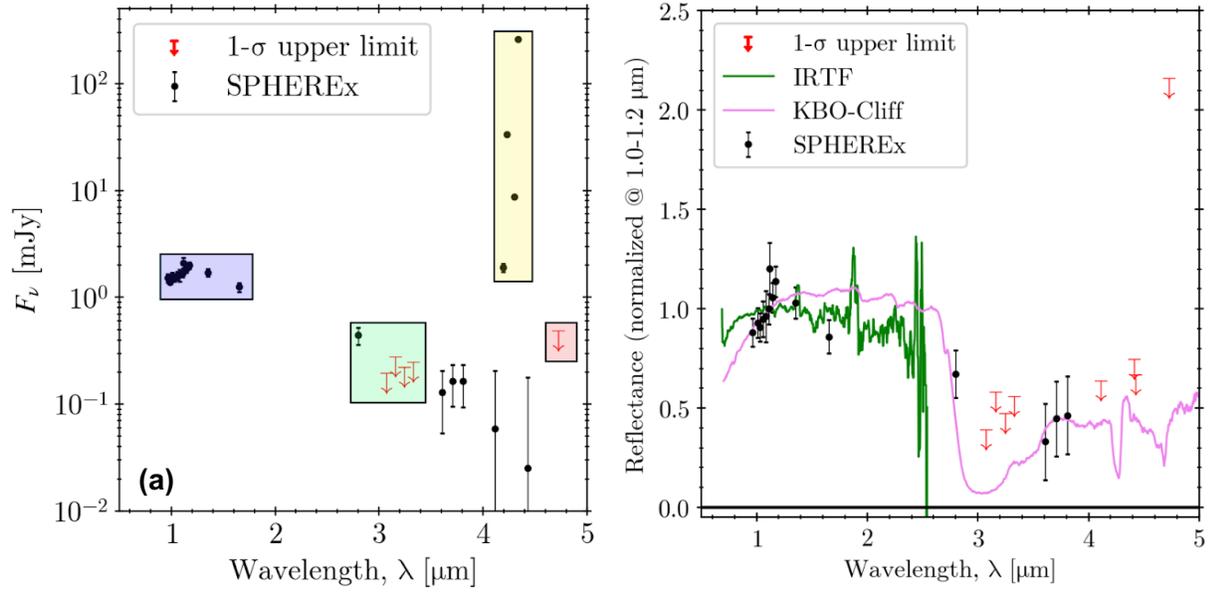

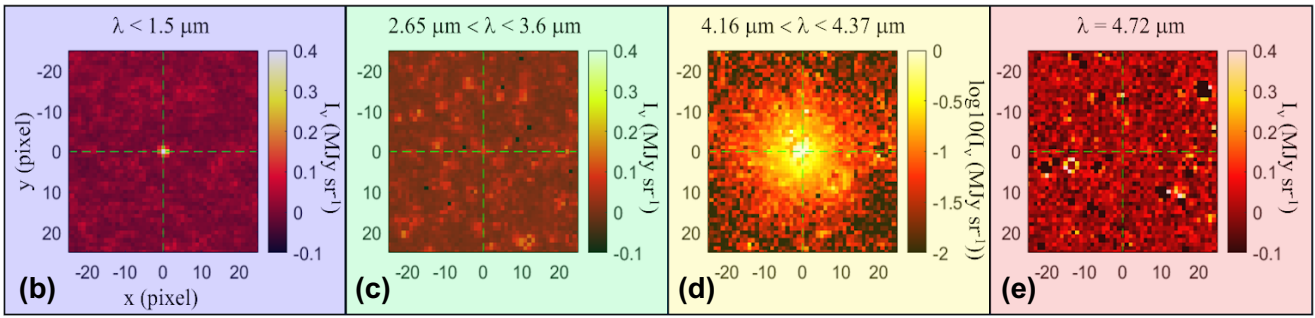

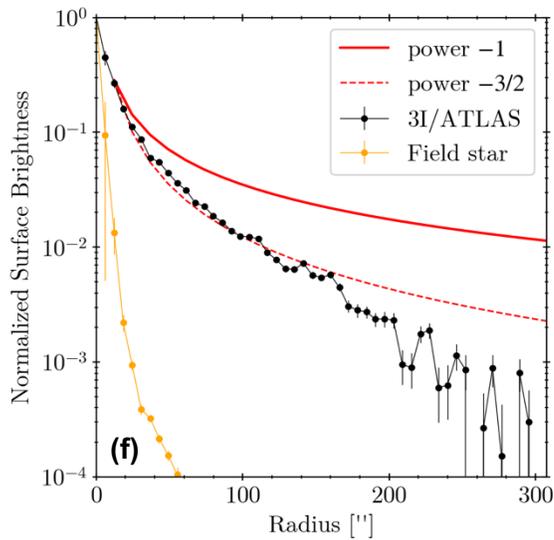

**Figure 1a : SPHEREx 0.75-5.0μm** 2-pixel radius aperture **spectrophotometry (*left*) and solar spectrum corrected reflectance (*right*) for 3I.** Red arrows denote possible significant background contamination due to known background sources. The arrows start at the measured flux+1σ. For the reflectance, the NASA-IRTF/SpeX 0.7-2.5μm spectrum (**green**) is overlaid. The two sets of measurements are consistent, and characterized by a slightly rising slope from 0.7-1.0μm, a flat regime from 1.0-1.5μm, a shallow absorption at 1.5 and 2.1μm, and a strong falloff starting at 2.4μm. The SPHEREx spectrophotometry stays low from 2.5-5.0μm, with a huge uptick at 4.3μm due to $CO_2$ gas emission. Also overlaid for comparison is the average "cliff" KBO spectrum of Pinilla-Alonso+2025 (**pink**). **Figure 1b: Stacked 1.0-1.5μm SPHEREx geometry- and Solar flux-corrected image of 3I.** Assuming the reflectance is flat across these wavelengths, this median stacking produces a deep image of the nucleus + dust scattered light continuum flux from 3I. There is no significant extension found in the stacked image beyond a stellar point source in a SPHEREx 6.15" pixel. **Figures 1c, d, and e: SPHEREx images centered at the $H_2O$/$CO_2$/CO coma gas wavelengths of 3.0/4.26/4.7μm, respectively.** 3I is undetected in $H_2O$ and CO, with 2-pixel radius preliminary conservative upper limits of 0.45 and 0.99mJy. By contrast, a bright $CO_2$ coma with central PSF flux = 33mJy that extends out to at least 30-pixel (348,000 km) was found. This observed minimum size for the $CO_2$ coma corresponds to ~700,000-second mean lifetime at $v_{gas}$ = 0.48 km/sec. The reported photodissociation lifetime for $CO_2$ at 1au is 135hrs = 486,000 seconds (Huebner+1992), which scales by $1/r_h^2$ to ~4,670,000 seconds at 3.2au. **Figure 1f: Azimuthally-averaged radial profile for the SPHEREx 3I $CO_2$ coma.** The observed falloff with projected distance is steeper than $1/\rho$, closer to $1/\rho^{3/2}$, which is consistent with a near-nucleus source of $CO_2$ coupled with photodissociation + charge exchange ionization processing of free-flying molecules in the coma.